\def\bar{\begin{eqnarray}}
\def\ear{\end{eqnarray}}

\def\eqi{\begin{linenomath*}\begin{equation}}
\def\eqf{\end{equation}\end{linenomath*}}
\def\eqia{\begin{eqnarray}}
\def\eqfa{\end{eqnarray}}

\def\oc2{$\mathcal{O}(c^{-2})$}

\documentclass{ws-procs975x65}

\begin{document}

\wstoc{The Lense-Thirring effect and the Pioneer anomaly: Solar
System tests}{L. Iorio}

\title{THE LENSE--THIRRING EFFECT AND THE PIONEER ANOMALY: SOLAR SYSTEM TESTS}

\author{LORENZO IORIO\footnote{Fellow of the Royal Astronomical Society}}

\address{
Viale Unit$\grave{ \it a}$ di Italia 68, 70125, Bari (BA),
Italy\\
\email{lorenzo.iorio@libero.it}}

\begin{abstract}
We report on a $\lessapprox 1\%$ test of the Lense-Thirring effect
with the Mars Global Surveyor  orbiter and on certain features of
motion of Uranus, Neptune and Pluto which contradict the
hypothesis that the Pioneer anomaly can be caused by some
gravitational mechanism.
\end{abstract}

\bodymatter

\section{The Lense-Thirring effect}
Up to now the Lense-Thirring effect\cite{1,2,3,4,5} has been only
tested in the terrestrial gravitational field with the LAGEOS
satellites\cite{6,7,8,9,10}. Although the relativistic predictions
are not in disagreement with the results of such tests, their
realistic accuracy has always been controversial\cite{8,11,12,13}.
%
Recent advances in planetary ephemerides$^{14}$ have made
meaningful to compare the relativistic predictions for the
Lense-Thirring effect of the Sun on the inner planets of the Solar
System\cite{15,16} to the least-squares determined extra-rates of
perihelion of such celestial bodies$^{14}$. There is no
contradiction between them; although the errors are still large so
that also a zero-effect cannot be ruled out, the hypothesis of the
existence of the solar gravitomagnetic field  is in better
agreement with the data$^{16}$.
In April 2004 the GP-B spacecraft\cite{17,18} has been launched to
measure the Schiff precession$^{19}$ of the spins of four
superconducting gyroscopes carried onboard: the expected accuracy
is $\approx 1\%$.
%
%
The space environment of Mars has recently yielded the opportunity
of performing another test$^{20}$ of the Lense-Thirring effect.
Almost six years of range and range--rate data of the Mars Global
Surveyor (MGS) orbiter, together with three years of data from
Odyssey, have been used in order to precisely determine many
global properties of Mars$^{21}$. As a by-product, also the orbit
of MGS has been very accurately reconstructed$^{21}$. The average
of the RMS overlap differences of the out-of-plane part of the MGS
orbit amounts to 1.613 m over an about 5-years time span (14
November 1999--14 January 2005). Neither the gravitomagnetic force
was included in the dynamical models used in the data reduction,
nor any empirical out-of-plane acceleration was fitted, so that
the RMS overlap differences entirely account for the martian
gravitomagnetic force. The average out-of-plane MGS Lense-Thirring
shift over the same time span amounts just to 1.610 m: a
discrepancy of $0.2\%$. The error has been evaluated as$^{20}$
$0.5\%$.
%
%
Let us, finally, note that the MGS test is based on a data
analysis done in a completely independent way with respect to the
author of Ref. $20$, without having gravitomagnetism in mind at
all.
\section{The Pioneer anomaly}
The Pioneer anomaly\cite{22,23,24}
 is an unexpected, almost constant and
uniform extra-acceleration $A_{\rm Pio}$ directed towards the Sun
of $(8.74\pm 1.33)\times 10^{-10}$ m s$^{-2}$ detected in the data
of both the Pioneer 10/11 probes after 20 AU.
%
It has attracted much interest because of the possibility that it
is a signal of some failure in the currently known laws of
gravitation$^{25,26}$.
%
%
%
If the Pioneer anomaly was of gravitational origin, it should then
fulfil the equivalence principle
and an extra-gravitational acceleration like $A_{\rm Pio}$ should
also affect the motion of any other object moving in the region in
which the Pioneer anomaly manifested itself. Uranus, Neptune and
Pluto  are ideal candidates to perform independent and clean tests
of the hypothesis that the Pioneer anomaly is due to some still
unexplained features of gravity. Indeed, their paths lie at the
edge of  the Pioneer anomaly region or entirely reside in it
because their semimajor axes   are 19.19 AU, 30.06 AU, and 39.48
AU, respectively, and their eccentricities  amount to $0.047$,
$0.008$ and $0.248$.
%
%
Under the action of $A_{\rm Pio }$, whatever physical mechanism
may cause it, their  perihelia would secularly precess at
unexpectedly large rates.
For Uranus, which is the only outer planet having completed a full
orbital revolution over the time span for which modern
observations are available, the anomalous perihelion rate is
$-83.58\pm 12.71$ arcseconds per century.
%
%
%
E.V. Pitjeva in processing almost one century of data with the
EPM2004 ephemerides$^{27}$ also determined extra-rates of the
perihelia of the inner$^{14}$ and outer$^{28}$ planets as fit-for
parameters of global solutions in which she contrasted, in a
least-square way, the observations
to their predicted values computed with a complete suite of
dynamical force models including all the known
features of motion. Thus, any unmodelled force as $A_{\rm Pio}$ is
entirely accounted for by the perihelia extra-rates. For the
perihelion of Uranus she preliminarily determined an extra-rate of
$+0.57\pm 1.30$ arcseconds per century. The quoted uncertainty is
just the mere formal, statistical error: the realistic one might
be up to $10-30$ times larger. Even if it was 50 times larger, the
presence of an unexpected precession as large as that predicted
for Uranus would be ruled out. It is unlikely that such a
conclusion will be substantially changed when further and
extensive re-analysis$^{29,30}$ of the entire Pioneer 10/11 data
set will be carried out since they will be focussed on what
happened well before 20 AU.
This result is consistent with the findings of Ref. $31$ in which
the time-dependent patterns of  $\alpha\cos\delta$ and $\delta$
induced by a Pioneer-like acceleration on Uranus, Neptune and
Pluto have been compared with the observational residuals
determined by Pitjeva$^{27}$ for the same quantities and the same
planets over a time span of about 90 years from 1913 (1914 for
Pluto) to 2003. While the former ones exhibited well defined
polynomial signatures of hundreds of arcseconds, the residuals did
not show any particular patterns, being almost uniform strips
constrained well within $\pm 5$ arcseconds over the data set time
span which includes the entire Pioneer 10/11 lifetimes.



\end{document}